\documentclass[aps,prb,showpacs,reprint, superscriptaddress]{revtex4-1}

\usepackage[T1]{fontenc}
\usepackage{graphicx}
\usepackage{amsmath}
\usepackage{bbm}
\usepackage{color}
\usepackage{float}
\input epsf

\begin{document}
\title{Even-parity spin-triplet pairing for orbitally degenerate
correlated
electrons by purely repulsive interactions}
\author{Micha\l \space Zegrodnik}
\email{michal.zegrodnik@gmail.com}
\affiliation{AGH University of Science and Technology,
Faculty of Physics and Applied Computer Science, Al. Mickiewicza 30,
30-059 Krak\'{o}w, Poland}
\author{J\"{o}rg B\"{u}nemann}
\email{buenemann@gmail.com}
\affiliation{Max-Planck Institute for Solid State Research,
Heisenbergstr. 1, D-70569 Stuttgart, Germany}
\author{Jozef Spa\l ek}
\email{ufspalek@if.uj.edu.pl}
\affiliation{AGH University of Science and Technology,
Faculty of Physics and Applied Computer Science, Al. Mickiewicza 30,
30-059 Krak\'{o}w, Poland}
\affiliation{Marian Smoluchowski Institute of Physics, 
Jagiellonian University, ul. Reymonta 4,
30-059 Krak\'{o}w, Poland}


\begin{abstract}
We demonstrate the stability of a spin-triplet paired s-wave
(with an admixture of extended s-wave)
state for the case of purely repulsive interactions in a degenerate two-band
Hubbard model. We further show that near half-filling the
considered kind of superconductivity can coexist with antiferromagnetism. The
calculations have been carried out with the use of the so-called
\textit{statistically consistent Gutzwiller approximation} for the case of a
square lattice. The absence of a stable paired state when analyzed in the
Hartree-Fock-BCS
approximation allows us to claim that the electron correlations in
conjunction with the Hund's rule exchange play
the
crucial role in stabilizing the spin-triplet superconducting state. A sizable
hybridization of the bands suppresses the paired state.
\end{abstract}


\maketitle

\textit{Introduction}.---Spin-triplet superconductivity was postulated to
occur in
Sr$_2$RuO$_4$\cite{Mackenzie2003}$^,$\cite{Rice1994}, in uranium
compounds\cite{Saxena2000}$^-$\cite{Tateiwa2001}, and in iron
pnictides\cite{Dai2008}$^,$\cite{Lee2008}. All these multi-band systems have
moderately
(Sr$_2$RuO$_4$ and the pnictides) or strongly correlated  (URhGe, UPt$_3$)
electrons, $d$ and
$f$, respectively. Earlier, the spin-triplet pairing has been used successfully
to describe the superfluidity of liquid $^3$He
\cite{Anderson1973}$^,$\cite{Anderson1978} and that of the neutron-star crust
\cite{Pines1965}. In the last two cases of fermionic systems, which are
considered as paramagnets with an enhanced susceptibility, a single-component (a
single-band)
Landau Fermi-liquid picture was taken as a starting point and the pairing of
the odd parity (p-wave) was
due
to the exchange of a paramagnon. Such an approach is limited to weak
correlations
and was also applied to weakly ferromagnetic superconducting
systems\cite{Fay1980} and to Sr$_2$RuO$_4$\cite{Mazin1997}.

In the correlated and orbitally degenerate systems the intraatomic
ferromagnetic
(Hund's rule) exchange
interaction of magnitude $J\sim0.1$eV, appears naturally in the extended
Hubbard model and is essential for
the description of ferromagnetism, for moderately and strongly correlated
electrons.
On the other
hand, its significance in the spin-triplet pairing has been emphasized in
general\cite{Spalek2001}$^-$\cite{Puetter2012}, as well as for both the
pnictides
\cite{Dai2008} and Sr$_2$RuO$_4$\cite{Takimoto2000}$^-$\cite{Koikegami}. In most
cases, the Hund's rule and other local Coulomb interactions are either
treated in
the Hartree-Fock approximation\cite{Zegrodnik2012} and/or semi-phenomenological
negative-$U$ intersite
attraction\cite{Annett2003} is introduced. A number of experimental results can
be successfully
interpreted in this manner, often assuming pairing with odd angular momentum,
though the situation in this respect is not yet completely
clear. In effect, it is very important to scrutinize a global
stability of the spin-triplet phase against an onset of either magnetism or
the coexistent
states within this canonical model of correlated electrons while treating both
the magnetism and the pairing in real space on equal footing.

We have recently analyzed a microscopic model with the Hund's-rule
induced
spin-triplet pairing, in both the Hartree-Fock\cite{Zegrodnik2012} and
the Gutzwiller approximation\cite{Zegrodnik2013}. In the Hartree-Fock-BCS limit,
the paired states
(often coexisting with magnetism) appear only in the limit $U'-J=U-3J<0$, where
$U'$ is
the intraatomic
interorbital magnitude of the Coulomb repulsion. This limit can be called as
that with attractive interactions. In the correlated Gutzwiller state and
under the same conditions, superconductivity, both pure and coexistent with
antiferromagnetism, is also
stable \cite{Zegrodnik2013}. The stability of superconducting phases comes not
as a surprise in this parameter regime, since it resembles a single band model
with negative U. In the course of this study, however, it became apparent to us
that the
spin-triplet paired
state can also become stable in the much more realistic regime of purely
repulsive interactions
$U'-J>0$, a typical situation for the correlated $3d$ and $4d$ electrons. The
purpose
of
this paper is to show that the s-wave (with a small
admixture of an extended s-wave) solution, i.e., with even parity, is stable and
therefore should be considered in the analysis of the spin-triplet
superconductivity in the orbitally degenerate and correlated systems. We would
like to underline that
this is a generic microscopic approach in
which the electronic correlations play a
decisive role in stabilizing the spin-triplet even-parity state. Namely, the
superconductivity induced by such pairing mechanism does not appear at all in
the Hartree-Fock-BCS type of approach.

\textit{Model}.---The starting Hamiltonian has the form of the extended Hubbard
model, i.e.,
\begin{equation}
\begin{split}
\hat{H}&=\sum_{ij(i\neq j)ll^{\prime}\sigma}t^{ll^{\prime}}_{ij}
\hat{c}_{il\sigma}^{\dag}\hat{c}_{jl^{\prime}\sigma}+U^{\prime}\sum_{i}\hat{
n } _ { i1}\hat{n}_{
i2}\\
&+U\sum_{il}\hat{n}_{
il\uparrow } \hat { n } _ {
il\downarrow}-J\sum_{ill^{\prime}(l\neq
l^{\prime})}\bigg(\mathbf{\hat{S}}_{il}\cdotp
\mathbf{\hat{S}}_{il^{\prime}}+\frac{1}{4}\hat{n}_{il}\hat{n}_{il^{\prime}}
\bigg)\;,
\label{eq:H_start}
\end{split}
\end{equation}
where $l=1,2$ labels the orbitals. The first term includes intraband ($l=l'$)
and interband (hybridization, $l\neq l'$) hopping terms, the second and third
represent the interorbital and intraorbital Coulomb repulsion, whereas the last
represents the full form of the Hund's rule exchange interaction. The
Hamiltonian (\ref{eq:H_start}) can be rewritten in a alternative form using the
real-space
representation for the pairing parts
\begin{equation}
\begin{split}
\hat{H}&=\sum_{ij(i\neq j)ll^{\prime}\sigma}t^{ll^{\prime}}_{ij}
\hat{c}_{il\sigma}^{\dag}\hat{c}_{jl^{\prime}\sigma}
+ U\sum_{il}\hat{n}_{il\uparrow}\hat{n}_{
il\downarrow}\\
&+(U^{\prime}+J)\sum_{i}\hat{B}_i^{\dagger}\hat{B}_i
+(U^{\prime}-J)\sum_ {im}\hat{A}^{\dagger}_{im} \hat{A}_{im}\;,
\label{eq:H_start2}
\end{split}
\end{equation}
where the spin-triplet $\hat{A}_{im}$ and spin-singlet $\hat{B}_{i}$ pairing
operators are defined as follows
\begin{equation}
\hat{A}^{\dagger}_{i,m}\equiv\left\{\begin{array}{cl}
\hat{c}^{\dagger}_{i1\uparrow}\hat{c}^{\dagger}_{i2\uparrow} & m=1\;,\\
\hat{c}^{\dagger}_{i1\downarrow}\hat{c}^{\dagger}_{i2\downarrow} & m=-1\;,\\
\frac{1}{\sqrt{2}}(\hat{c}^{\dagger}_{i1\uparrow}\hat{c}^{\dagger}_{i2\downarrow
} +\hat{c}^ { \dagger
}_{i1\downarrow}\hat{c}^{\dagger}_{i2\uparrow}) & m=0\;,\\
\end{array}\right.
\label{eq: A_op}
\end{equation}
\begin{equation}
 \hat{B}^{\dagger}_i\equiv\frac{1}{\sqrt{2}}(\hat{c}^{\dagger}_{i1\uparrow}\hat{c}^{
\dagger } _ {
i2\downarrow }
-\hat{c}^{\dagger}_{i1\downarrow}\hat{c}^{\dagger}_{i2\uparrow})\;.
\end{equation}
As one can see, for $U'>J$ the interaction energy
that
corresponds to the creation of a single pair in either spin-triplet
or spin-singlet states on
a atomic site, is positive. For an
orbitally degenerate case, where the standard hierarchy of couplings is
$U>U'>J$, the interorbital local spin-triplet type of pairing, if any, may
be favored over the singlet one. The
factor favoring the triplet over the singlet pairing is the
Hund's rule exchange, but as we show, the electronic correlations
are equally important to stabilize the paired state globally.

\textit{Method}.---As said above, electronic correlations turn out to be crucial in this
system. To include them in our study we use the modified Gutzwiller
approximation. In this method, one assumes that the correlated
state $|\Psi_G\rangle$ of the system can be expressed in the following manner
\begin{equation}
|\Psi_G \rangle=\hat{P}_G|\Psi_0 \rangle\;,
\end{equation}
where $|\Psi_0\rangle$ is the normalized non-correlated state to be defined
below, whereas
$\hat{P}_G$ is the Gutzwiller correlator, which we have selected in the
form
\begin{equation}
 \hat{P}_G=\prod_i\hat{P}_{G|i}\equiv\prod_{i}\sum_{I,I^{\prime}}\lambda_{I,I^{
\prime } } |I\rangle_ { ii }
 \langle I^{\prime}|\;.
\label{eq:correlator}
\end{equation}
Here, $\{|I\rangle\}$ is a basis of the local (atomic) Hilbert space (16 states)
and
$\lambda_{I,I'}$ are variational parameters, which we assume to be
real. In the subsequent discussion, we write the expectation values with
respect to $|\Psi_0\rangle$ as $\langle\hat{O}\rangle_0\equiv\langle
\Psi_0|\hat{O}|\Psi_0\rangle$,
while the expectation values with respect to $|\Psi_G\rangle$ will be denoted by
\begin{equation}
 \langle\hat{O}
\rangle_G\equiv\frac{\langle\Psi_G|\hat{O}|\Psi_G\rangle}{
\langle\Psi_G|\Psi_G\rangle
}=\frac{\langle\Psi_0|\hat{P}_G\hat{O}\hat{P}_G|\Psi_0\rangle}{
\langle\Psi_0|\hat{P}_G^2|\Psi_0\rangle
}\;.
\end{equation}
We focus on the pure superconducting phase of type A for which
$\langle\hat{A}_{i,1}\rangle_G=\langle\hat{A}_{i,-1}\rangle_G\neq 0$ and
$\langle\hat{A}_{i,0}\rangle_G\equiv0$. This is because one would expect that
the
equal spin state (ESP) is favored by the local ferromagnetic exchange. Note that the
expectation
values in the correlated state, $|\Psi_G\rangle$ of the respective pairing
operators are nonzero only if the corresponding expectation values in
the noncorrelated state $|\Psi_0\rangle$ are also nonzero.
For simplicity, we assume that $t^{11}=t^{22}\equiv t$ and
$t^{12}=t^{21}\equiv
t'$ for the nearest neighbors. The expectation value of the grand Hamiltonian
$\hat{K}=\hat{H}-\mu\hat{N}$ in the correlated state has been derived in the
limit of
infinite dimensions by a diagrammatic
approach\cite{Bunemann2005} and has the form
\begin{equation}
\begin{split}
 \langle\hat{K}\rangle_G&=\sum_{ijl\sigma}Q\;t_{ij}\langle\hat{c}^{\dagger}_{
il\sigma } \hat { c}_{jl\sigma}\rangle_0 +
\sum_{ijll'\sigma}Q\;t_{ij}'\langle\hat{c}^{\dagger}_{il\sigma}
\hat { c}_{jl'\sigma}\rangle_0\\
&+\sum_{ij\sigma}\tilde{Q}\;t_{ij}(\langle\hat{c}^{\dagger}_{
i1\sigma } \hat{c}^{\dagger}_{j2\sigma}\rangle_0+\langle\hat{c}_{
j2\sigma } \hat{c}_{i1\sigma}\rangle_0)\\
&+L\sum_{I,I'}\bar{E}_{I,I'}\langle\hat{m}_{I,I'}\rangle_0-
\mu\sum_{il\sigma}q^s_{l\sigma}\langle\hat{n}_{il\sigma}\rangle_0\;,\\
\end{split}
\label{eq:H_expe}
\end{equation}
where $Q$ and $\tilde{Q}$ are the renormalization factors, L is the
number of atomic sites, $\mu$ refers to the chemical
potential, $q_{l\sigma}^s=\langle\hat{n}_{il\sigma}\rangle_G/\langle\hat{n}_{
il\sigma}\rangle_0$, and $\hat{m}_{I,I'}\equiv|I\rangle\langle I'|$. The
factors $Q$ and
$\tilde{Q}$, as well as $\bar{E}_{I,I'}$, can be expressed
with the use of the variational parameters $\lambda_{I,I'}$, the local single
particle density matrix elements
$\langle\hat{c}^{\alpha}_{il\sigma}\hat{c}^{\alpha'}_{il'\sigma'}\rangle_0$, and the
matrix elements of the atomic part of (\ref{eq:H_start})
represented in the local basis, $\langle I|\hat{H}^{at}|I' \rangle$. Here
$\hat{c}^{\alpha}_{il\sigma}$ are either creation or anihilation operators. The
expression for $\langle\hat{K}\rangle_G$
can be rewritten as the expectation value of the effective
single-particle
Hamiltonian
$\hat{K}_{GA}$, evaluated with respect to $|\Psi_0\rangle$, i.e.,
\begin{equation}
\begin{split}
 \hat{K}_{GA}&=\sum_{ijl\sigma}Q\;t_{ij}\hat{c}^{\dagger}_{
il\sigma } \hat { c}_{jl\sigma} +
\sum_{ijll'\sigma}Q\;t_{ij}'\hat{c}^{\dagger}_{il\sigma}
\hat { c}_{jl'\sigma}\\
&+\sum_{ij\sigma}\tilde{Q}\;t_{ij}(\hat{c}^{\dagger}_{
i1\sigma } \hat{c}^{\dagger}_{j2\sigma}+\hat{c}_{
j2\sigma } \hat{c}_{i1\sigma})\\
&+L\sum_{I,I'}\bar{E}_{I,I'}\langle\hat{m}_{I,I'}\rangle_0-
\mu\sum_{il\sigma}q^s_{l\sigma}\hat{n}_{il\sigma}\;.\\
\end{split}
\label{eq:H_GA}
\end{equation}
The first three terms of (\ref{eq:H_GA}) originate from the single particle part
of (\ref{eq:H_start2}), while the fourth originates from its interaction
part. It can be seen that the intraatomic part has been taken as its average,
in accordance with the general philosophy of the Gutzwiller approach. Again, the
$Q$ and $\tilde{Q}$ factors are the renormalization factors of
the
respective dynamic processes. The first two refer to the
narrowing of the quasiparticle bands, whereas the
$\tilde{Q}$ parameter corresponds to the intersite pairing
amplitude. It should be emphasized that in our
initial Hamiltonian (\ref{eq:H_start}) there are no intersite interaction
terms and so the intersite pairing that is present in (\ref{eq:H_GA}) is due to
correlations (a non-BCS factor). Also, the factor $\tilde{Q}$ is nonzero only
when the local
expectation values $\langle\hat{A}_{i,\pm 1} \rangle_G$ (and the corresponding
$\langle\hat{A}_{i,\pm 1} \rangle_0$)
are also nonzero. As a result, the intersite pairing appears concomitantly with
the intrasite one. 

In the statistically consistent Gutzwiller approach
(SGA)\cite{Jedrak2010}$^,$\cite{Kaczmarczyk2011} the
mean fields are treated as variational parameters, with respect to which the
free energy of the system is
minimized. Hence, in order
to assure that the self-consistent and the variational procedures yield the same
results, additional
constraints have to be introduced with the help of the Lagrange-multiplier
method. This leads to supplementary terms in the effective Hamiltonian so that
now it takes the form 
\begin{equation}
\begin{split}
 \hat{K}_{\lambda}&=\hat{K}_{GA}-\lambda_n\big(\sum_{il\sigma}q^s_{
l\sigma}\hat{
n}_{il\sigma}-L\langle \hat{n}\rangle_G\big)\\
&-\sum_{m=\pm 1}\big[\lambda_{m}\big(\sum_{i}\hat{A}_{im}-L\langle\hat{A}_{im}\rangle_0\big)
+H.C.\big]\;,\\
\end{split}
\label{eq:H_lambda}
\end{equation}
where the Lagrange multipliers $\lambda_m$ and $\lambda_n$ are introduced to
assure
that the averages $\langle\hat{A}_{im}\rangle$ and
$\langle\hat{n}\rangle$ calculated either from the corresponding
self-consistent equations or variationally, coincide with each other. One
should also note that it is natural to fix $\langle\hat{n}\rangle_G$
instead of
$\langle\hat{n}\rangle_0$ during the
minimization procedure. This is the reason why we put
the term $-\mu\hat{N}$ already at the beginning of out derivation. The values
of the mean fields, the
variational
parameters, the and the Lagrange multipliers, are all found by minimizing the
free energy
functional $\hat{F}_{\lambda}$ that is derived with the help of the effective
Hamiltonian $\hat{K}_{\lambda}$ in a standard statistical-mechanical manner. For
the
considered two-band model there can be up to 256 variational parameters
$\lambda_{I,I'}$. Fortunately, for symmetry reasons, one can reduce their
number significantly. It should also be noted that not all of the parameters are
independent, as certain constraints
have to be obeyed \cite{Zegrodnik2013}$^,$\cite{Bunemann2005}. In
effect, we have to minimize only 16 variables in this pure superconducting
state of type A.

From
Eqs.(\ref{eq:H_GA}) and
(\ref{eq:H_lambda}) it can be seen that the Lagrange multipliers $\lambda_{m}$
have
an interpretation of the intrasite gap parameters, while the symmetry of the
intersite gap parameter is fully determined by the bare band dispersion
relation. By assuming the dispersion relation for a square lattice with nonzero
hopping $t$ between nearest neighbors only
\begin{equation}
 \epsilon_{\mathbf{k}}=-2t(\cos k_x +\cos k_y)\;,
\label{eq:dispersion}
\end{equation}
one obtains the following form of the gap parameter
\begin{equation}
 \Delta_{\mathbf{k}}=\Delta^{(0)}+\Delta^{(1)}(\cos{k_x}+\cos{k_y})\;,
\end{equation}
where $\Delta^{(0)}\equiv\lambda_{1}=\lambda_{-1}$ (as we are considering an
ESP
state) while $\Delta^{(1)}\equiv2\tilde{Q}t$ is the intersite pairing amplitude. In
this manner, we have obtained a mixture of the s-wave and the extended s-wave
pairing
symmetry. 

In order to check if the stable spin-triplet paired
phases can indeed appear in the repulsive-interaction regime, we have performed first the calculations taking into
account only the
intrasite pairing for the following selection of phases: type A
superconducting ({\bf A}), pure ferromagnetic ({\bf FM}), paramagnetic ({\bf
NS}),
superconducting coexisting with antiferromagnetism ({\bf SC+AF}), and
pure antiferromagnetic ({\bf AF}). The antiferromagnetic ordering
considered by
us has a simple two-sublattice form. 
We have also considered the so-called A1
superconducting phase ($\langle
\hat{A}_1\rangle_G\neq 0$ and
$\langle
\hat{A}_{-1}\rangle_G=\langle
\hat{A}_0\rangle_G\equiv0$) coexisting with ferromagnetism. However, this phase
turned
out not to be stable for the whole range of model parameters examined.
Therefore, it is not included in the subsequent discussion. Detailed
information concerning the above phases can be found in
\cite{Zegrodnik2013}, where we have analyzed the intrasite paired states in
the regime of attractive interaction, i.e., for $U'-J<0$.

\textit{Results}.---The calculations have been performed assuming that the
hybridization matrix element has the form
$\epsilon_{12\mathbf{k}}\equiv \beta_h\epsilon_{\mathbf{k}}$,
where $\beta_h\in[0,1]$, specifies the interband hybridization
strength. The interorbital Coulomb repulsion constant $U'$ was set to
$U'=U-2J$. All the energies have been
normalized to the bare band-width, $W=8|t|$, and the presented results were
obtained for $k_BT/W=10^{-4}$ emulating the $T=0$ state.

\begin{figure}[b!]
\centering
\epsfxsize=85mm 
{\epsfbox[190 300 537 600]{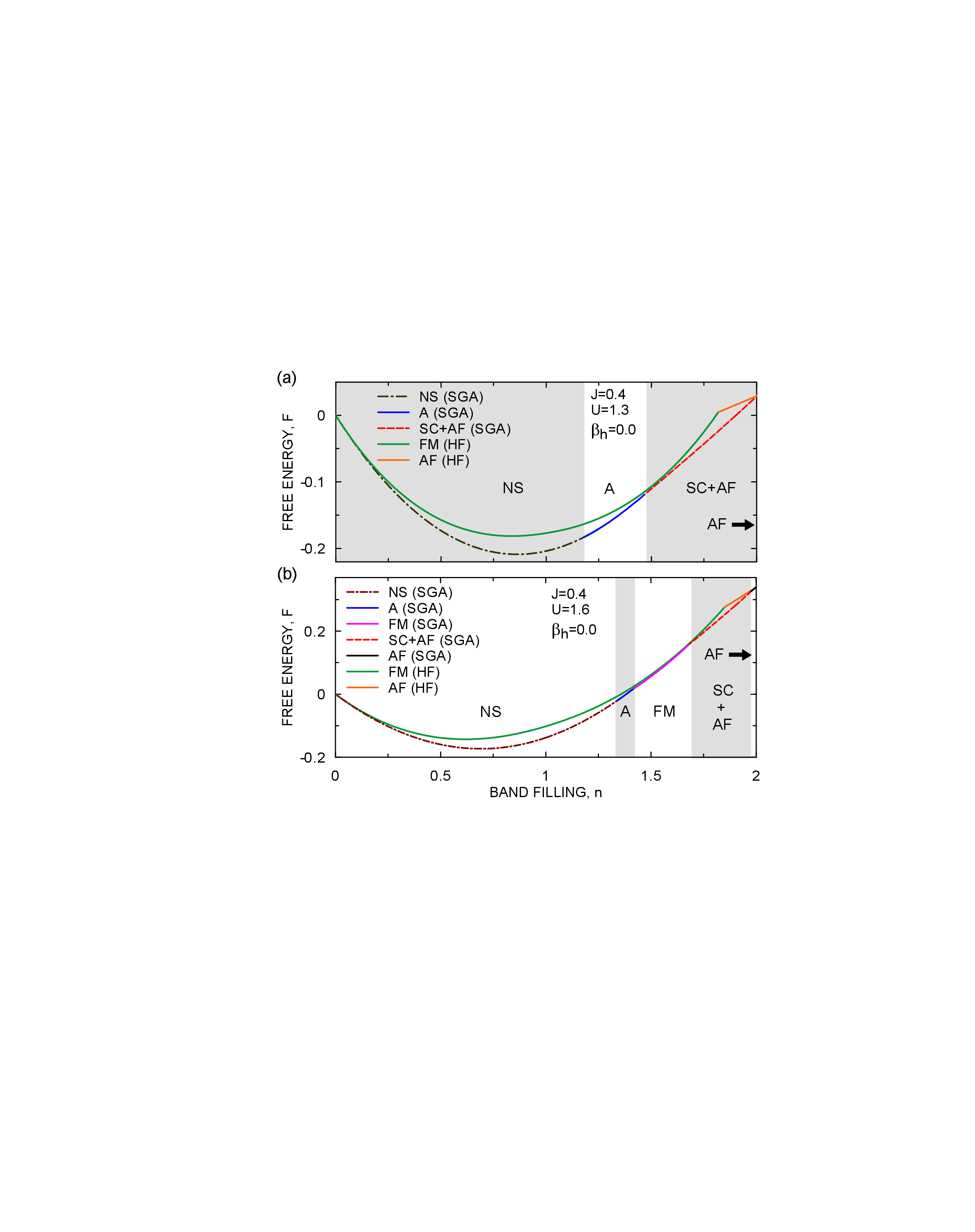}}
\caption{(Color online) Ground-state energy of stable phases as a function of
the band
filling for the case when only the intrasite pairing is included (i.e., for
$\Delta^{(1)}\equiv0$). For
comparison, plots obtained in the H-F approximation
are also shown. The shaded regions mark the stability of corresponding
phases according to the SGA method. Pure AF state is stable for $n=2$ (marked
by arrow).}
\label{fig:1}
\end{figure}

In Fig. \ref{fig:1} we show that the
superconducting phases, both pure and coexisting with antiferromagnetism, are
stable for purely repulsive interactions regime ($U'-J>0$). With the increasing Coulomb repulsion $U$, the regions of stability of
the paired phases are becoming
narrower. Note that the Hartree-Fock calculations lead only to the stability of
magnetically ordered
phases in this regime. The appearance of
the paired states is therefore a genuine many-particle effect which is caused
by the electronic correlations and taken into
account in the SGA method.

\begin{figure}[h!]
\centering
\epsfxsize=80mm 
{\epsfbox[97 453 509 653]{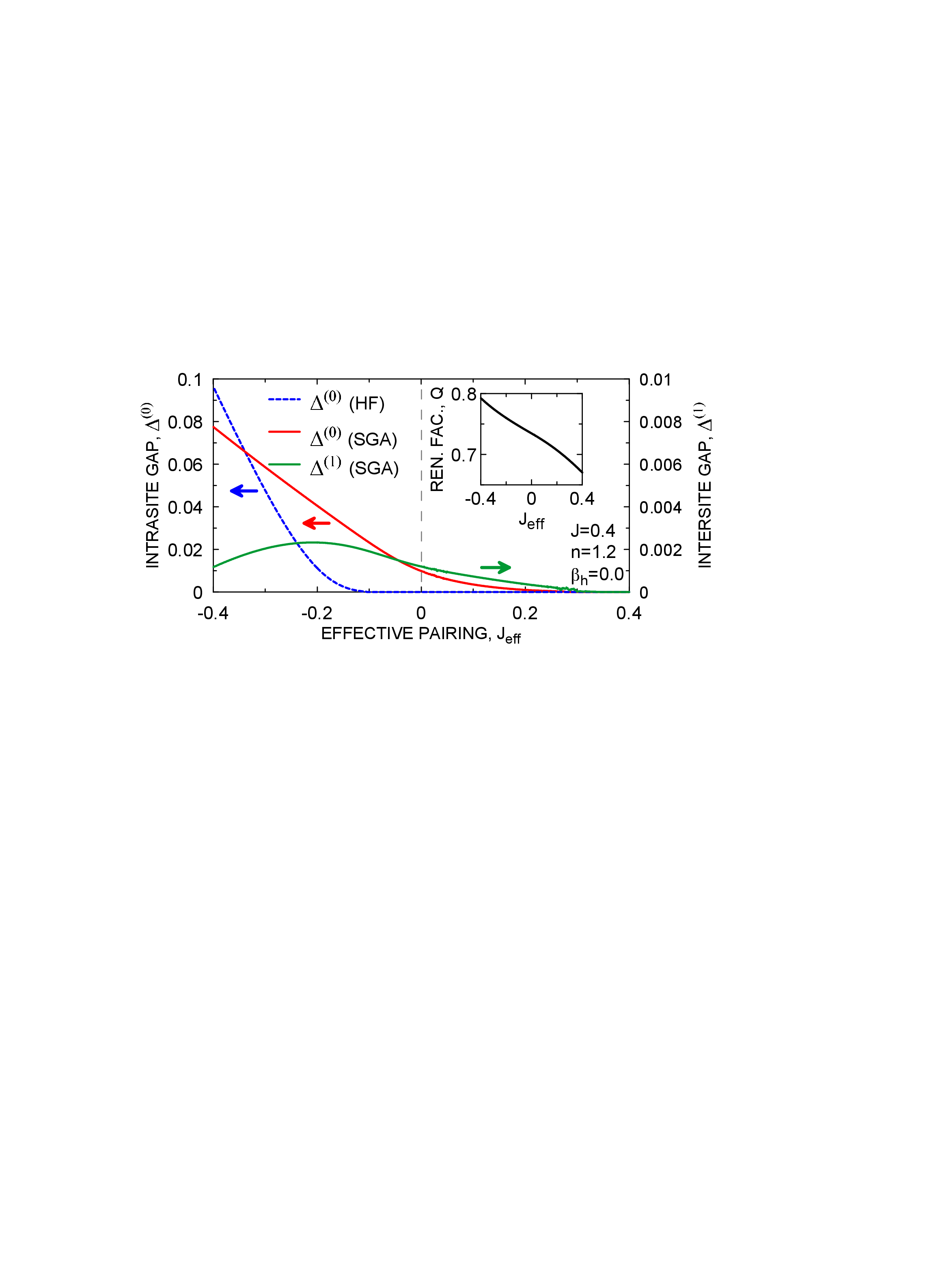}}
\caption{(Color online) The intrasite (left scale) and the intersite (right
scale) gap components as a functionof the effective coupling constant
$J_{\mbox{eff}}=U'-J$. For comparison, we
provide also the results obtained in the Hartree-Fock approximation.
Additionally, the band renormalization factor is shown in the inset. Note that $\Delta^{(1)}=\tilde{Q}/4$.}
\label{fig:2}
\end{figure}

Next, we discuss the superconducting A phase with inclusion of the intersite
part of the pairing. In Fig. \ref{fig:2} we plot the superconducting
gap components as a function of the effective
pairing constant $J_{\mathrm{eff}}\equiv U'-J$. As the value of the
$J_{\mathrm{eff}}$ parameter changes sign to positive, the intra-site
interaction corresponding to the spin-triplet-pair creation on a single atomic
site changes from attractive to repulsive. As one could
expect,
according to the
Hartree-Fock-BCS results, the intrasite gap parameter vanishes before
$J_{\mathrm{eff}}$
reaches zero and the intersite pairing does not appear. The
situation is different in the SGA. Namely, the paired
solution survives for
$J_{\mathrm{eff}}>0$ and the pairing has both the intra- and the inter-site
components.
However, the $\Delta^{(1)}$ parameter is an order of magnitude smaller than
$\Delta^{(0)}$. The phase A has a lower value of energy
than the normal phase for the whole range of $J_{\mathrm{eff}}$ presented in
Fig. \ref{fig:2}. Exemplary values of the order parameters, the
renormalization factors, and the free energy for $T\rightarrow 0$, are all
listed in Table \ref{table: values}. 
\begin{table}[h]
\centering
\caption{Representative values of the gap parameters, the renormalization
factors
and the
free energies for $J=0.4$, $n=1.2$ and $\beta_h=0.0$, for three different values
of the effective pairing constant,
$J_{\mbox{eff}}$. For comparison, we have
provided the values of the
renormalization factor and the free energy for the superconducting
phase of type A and the normal phase, NS. The subscripts refer to these two
phases. The numerical accuracy is better that the
last digit specified.}
 \begin{tabular}{||c||c|c|c|c|c|c||}
  \hline
   $J_{\mbox{eff}}$ & $\Delta^{(0)}$ & $\Delta^{(1)}$ & $Q_{A}$ & $Q_{NS}$ &
$F_{A}$ & $F_{NS}$   \\
  \hline
   $ -0.1 $ & 0.02325 & 0.00191 & 0.74669 & 0.74401 & -0.255481 & -0.255067 \\
   $ 0.1 $  & 0.00357 & 0.00073 & 0.72164 & 0.72157 & -0.179725 & -0.179705 \\
   $ 0.15 $ & 0.00200 & 0.00054 & 0.71454 & 0.71453 & -0.161874 & -0.161867 \\
  \hline
 \end{tabular}
 \label{table: values}
\end{table}

\begin{figure}[h!]
\centering
\epsfxsize=80mm 
{\epsfbox[71 380 613 661]{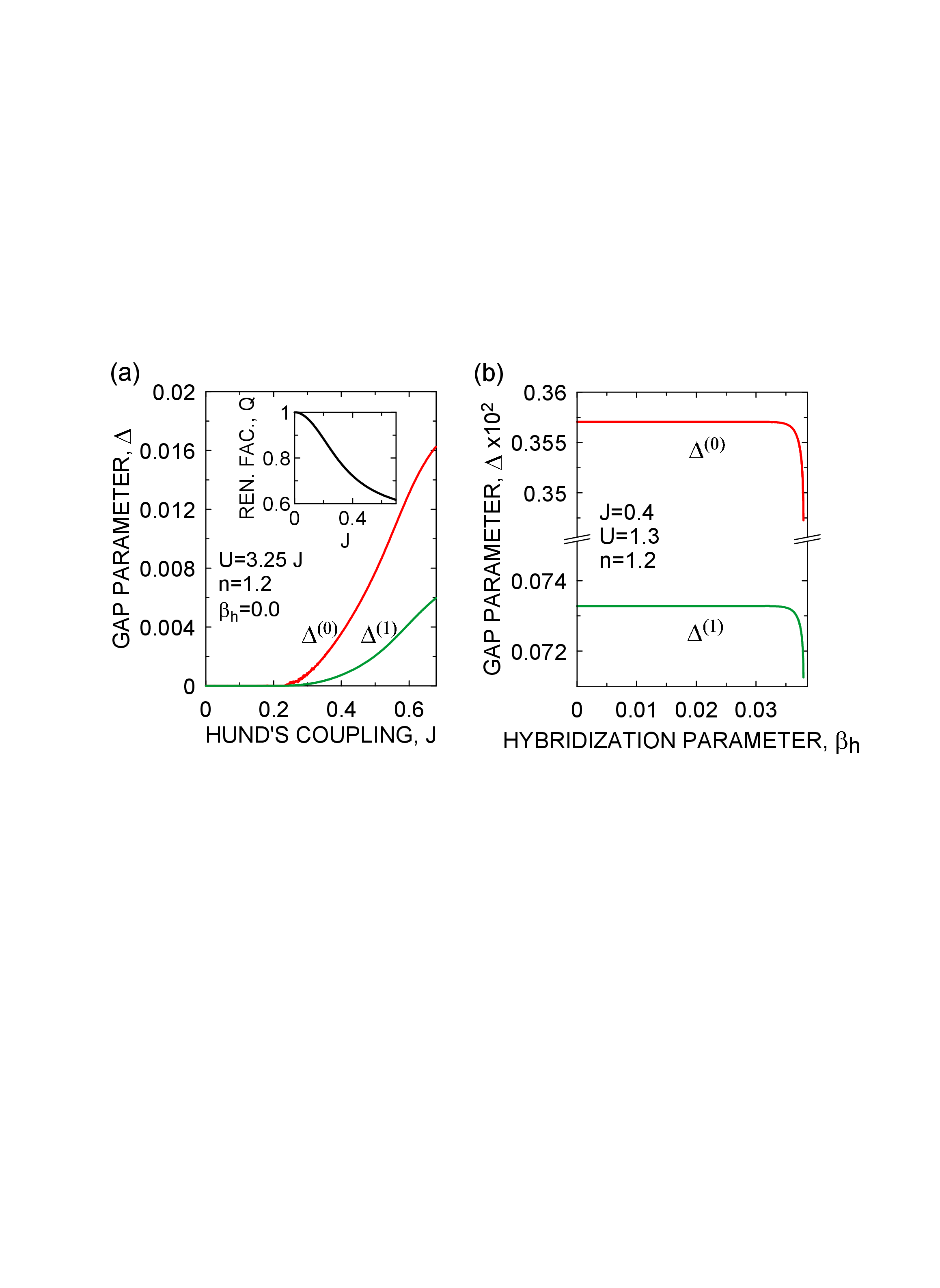}}
\caption{(Color online) Intrasite ($\Delta^{(0)}$) and intersite
($\Delta^{(1)}$) gap parameters as a function of the Hund's exchange integral
(a) and the hybridization parameter (b). In the
inset of (a) the $J$ dependence of the band narrowing factor is shown.}
\label{fig:3}
\end{figure}

In Fig. \ref{fig:3}a we plot the $J$ dependences of the gap parameters for
$J_{\mbox{\small{eff}}}=0.1$. For larger values of $J$ the
difference in magnitude between the intra- and the inter-site contributions to
the pairing is not that large. The
influence of the hybridization on the considered type of superconductivity is
shown in Fig. \ref{fig:3}b. The superconducting gaps are not affected by the
increase of the $\beta_h$ parameter up to the critical value
$\beta_h^C\approx0.0379$
at which both of
them suddenly drop to zero. Therefore, a sizable hybridization is detrimental to
the spin-triplet pairing and the effect is strong. It means that this type of
pairing suppresses the energy gain due the
interorbital hopping and hence is possible only for weakly hybridized
systems, where the condensation energy is dominant,
$2(\Delta^{(0)})^2/J\gtrsim\beta_h/8$.

\textit{Conclusions}.---By using the SGA approach, we have shown that
the
intrasite spin-triplet paired states, both pure (A type) and coexistent with
antiferromagnetism (SC+AF phase) can become stable in
the orbitally degenerate Hubbard model, in the limit of purely repulsive
interactions ($U'-J>0$). The coexistent SC+AF phase is possible for the systems
close to the half filling (the case of pnictides), whereas the pure A phase
appears when $n\approx 1.2$ for doubly or when $n\approx 1.8$ for triply
degenerate band
which corresponds roughly to the case of Sr$_2$RuO$_4$ in the hole language. We
have also analyzed
the intersite pairing
appearance for the considered regime of microscopic parameters. One can say
that both the Hund's rule and the correlations induced
change of band energy contribute to the spin-triplet pairing
mechanism; they correspond to the BCS (potential energy gain) and the non-BCS
(kinetic energy gain) factors stabilizing the paired state. The intersite
(extended
s-wave) part of the pairing is related to the intrasite (s-wave) one. This can
be seen from Figs.
\ref{fig:2} and \ref{fig:3}a,  where
$\Delta^{(0)}$ and $\Delta^{(1)}$ reach zero for the same values of model
parameters. The hybridization is detrimental to the superconducting A-phase
stability when the spin-triplet pairing condensation energy becomes smaller than
the
Pauli-principle-allowed kinetic
energy gain. We believe that the combined Hund's-rule and
correlation-induced
pairing presented here in the canonical model for the description of itinerant
magnetism opens up new
possibilities to study the spin-triplet superconductivity and its
coexistence with magnetic ordering in realistic multi-band systems. Within this
approach the spin-fluctuation contribution is of higher order.

M.Z. has been partly supported by the EU Human Capital Operation
Program, Polish
Project No. POKL.04.0101-00-434/08-00. This work has been partly supported by
the Foundation for Polish Science (FNP) within project TEAM and partly
by the National Science Center (NCN), through scheme MAESTRO, Grant No.
DEC-2012/04/A/ST3/00342. We are also grateful to Karol I. Wysoki\'nski for helpful
discussions.



\begin{thebibliography}{99}

\bibitem{Mackenzie2003}
A. P. Mackenzie and Y. Maeno, Rev. Mod. Phys. {\bf 75}, 657 (2003).

\bibitem{Rice1994}
T. M. Rice and M. Sigrist, J. Phys.: Condens Matter {\bf 7}, L643 (1994).

\bibitem{Saxena2000}
S. S. Saxena, P. Agarwal, K. Ahilan, F. M. Grosche, R.
K. W. Haselwimmer, M. J.
Steiner, E. Pugh, I. R. Walker, S. R. Julian, P. Monthoux, G. G. Lonzarich, A.
Huxley, I. Sheikin, D. Braithwaite and J. Flouquet, Nature {\bf 406}, 587
(2000).

\bibitem{Huxley2001}
A. Huxley, I. Sheikin, E. Ressouche, N. Kemovanois, D. Braithwaite, R.
Calemczuk, J. Flouquet, Phys. Rev. B {\bf 63}, 144519 (2001).

\bibitem{Tateiwa2001}
N. Tateiwa, T. C. Kobayashi, K. Hanazono, K. Amaya, Y.
Haga, R. Settai and Y.
Onuki, J. Phys.: Condens. Matter {\bf 13}, 117 (2001).

\bibitem{Dai2008}
X. Dai, Z. Fang, Y. Zhou, and F-C. Zhang, Phys. Rev. Lett {\bf 101}, 057008
(2008).

\bibitem{Lee2008}
P. A. Lee and X-G. Wen, Phys. Rev. B {\bf 78}, 144517 (2008).

\bibitem{Anderson1973}
P. W. Anderson and W. F. Brinkman, Phys. Rev. Lett. {\bf 30}, 1108 (1973).

\bibitem{Anderson1978}
P. W. Anderson and W. F. Brinkman in
\textit{Physics of Liquid and Solid Helium}, edited by K. H. Bennemann, and J.
B. Ketterson (J. Wiley \& Sons, New
York, 1978) Part II, pp. 177-286.

\bibitem{Pines1965}
D. Pines and A. Alpar, Nature {\bf 316}, 27 (1985).

\bibitem{Fay1980}
D. Fay and J. Appel, Phys. Rev. B {\bf 22}, 3173 (1980).

\bibitem{Mazin1997}
I. I. Mazin and D. I. Singh, Phys. Rev. Lett. {\bf 79}, 733 (1997).

\bibitem{Spalek2001}
J. Spa\l ek, Phys. Rev. B {\bf 63}, 104513 (2001).

\bibitem{Klejnberg1999}
A. Klejnberg and J. Spa\l ek, J. Phys. C: Condens. Matter {\bf 11}, 6553 (1999).

\bibitem{Han2004}
J. E. Han, Phys. Rev. B {\bf 70}, 054513 (2004).

\bibitem{Sano2003}
K. Sano and Y. Ono, J. Phys. Soc. Jpn {\bf 72}, 1847 (2003).

\bibitem{Puetter2012}
C. M. Puetter and H-Y. Kee, Eur. Phys. Lett. {\bf 98}, 27010 (2012).

\bibitem{Takimoto2000}
T. Takimoto, Phys. Rev. B {\bf 62}, R14641 (2000).

\bibitem{Koikegami}
S. Koikegami, Y. Yoshida, and T. Yanagisawa, Phys. Rev. B {\bf 67}, 134517
(2003).

\bibitem{Zegrodnik2012}
M. Zegrodnik and J. Spa\l ek, Phys. Rev. B {\bf 86} 014505 (2012).

\bibitem{Annett2003}
J. F. Annett, B. L. Gy\"orffy, G. Litak, and K. I. Wysoki\'nski, Eur. Phys. J.
B {\bf 36}, 301 (2003);\newline
J. F. Annett, B. L. Gy\"orffy, and K. I. Wysoki\'nski, New J. Phys. {\bf 11},
055063 (2009);\newline
K. I. Wysoki\'nski, J. F. Annet, and B. L. Gy\"orffy,
Phys. Rev. Lett. {\bf 108}, 1077004 (2012).

\bibitem{Zegrodnik2013}
M. Zegrodnik, J. Spa\l ek, and J. B\"unemann, arxiv:
1304.4478, unpublished

\bibitem{Bunemann2005}
J. B\"{u}nemann, F. Gebhard, T. Ohm, S. Weiser, and W. Weber in
\textit{Frontiers in Magnetic Materials}
(Springer, Berlin 2005).

\bibitem{Jedrak2010}
J. J\k{e}drak and J. Spa\l ek, Phys. Rev. B {\bf 81} 073108 (2010).

\bibitem{Kaczmarczyk2011}                                          
J. Kaczmarczyk and J. Spa\l ek, Phys. Rev. B {\bf 84} 125140 (2011).

\end{thebibliography}
\end{document}